\def \SAIT #1 #2 {{\em Mem.\ Soc.\ Astron.\ It.\/} {\bf #1}, #2}
\def \MESS #1 #2 {{\em The Messenger\/} {\bf #1}, #2}
\def \ASTRNACH #1 #2 {{\em Astron. Nach.\/} {\bf #1}, #2}
\def \AAP #1 #2 {{\em Astron. Astrophys.\/} {\bf #1}, #2}
\def \AAL #1 #2 {{\em Astron. Astrophys. Lett.\/} {\bf #1}, L#2}
\def \AAR #1 #2 {{\em Astron. Astrophys. Rev.\/} {\bf #1}, #2}
\def \AAS #1 #2 {{\em Astron. Astrophys. Suppl. Ser.\/} {\bf #1}, #2}
\def \AJ #1 #2 {{\em Astron. J.\/} {\bf #1}, #2}
\def \ANNREV #1 #2 {{\em Ann. Rev. Astron. Astrophys.\/} {\bf #1}, #2}
\def \APJ #1 #2 {{\em Astrophys. J.\/} {\bf #1}, #2}
\def \APJL #1 #2 {{\em Astrophys. J. Lett.\/} {\bf #1}, L#2}
\def \APJS #1 #2 {{\em Astrophys. J. Suppl.\/} {\bf #1}, #2}
\def \APSS #1 #2 {{\em Astrophys. Space Sci.\/} {\bf #1}, #2}
\def \ASR #1 #2 {{\em Adv. Space Res.\/} {\bf #1}, #2}
\def \BAIC #1 #2 {{\em Bull. Astron. Inst. Czechosl.\/} {\bf #1}, #2}
\def \JSQRT #1 #2 {{\em J. Quant. Spectrosc. Radiat. Transfer\/} {\bf #1}, #2}
\def \MN #1 #2 {{\em Mon. Not. R. Astr. Soc.\/} {\bf #1}, #2}
\def \MEM #1 #2 {{\em Mem. R. Astr. Soc.\/} {\bf #1}, #2}
\def \PLR #1 #2 {{\em Phys. Lett. Rev.\/} {\bf #1}, #2}
\def \PASJ #1 #2 {{\em Publ. Astron. Soc. Japan\/} {\bf #1}, #2}
\def \PASP #1 #2 {{\em Publ. Astr. Soc. Pacific\/} {\bf #1}, #2}
\def \NAT #1 #2 {{\em Nature\/} {\bf #1}, #2}
\def\bge{\begin{equation}}
\def\ede{\end{equation}}

\documentstyle[twoside]{memsait}
\input epsf.sty
\begin{opening}
\title{The Role of Hydrogen on Mass Loss from Proto-Globular Clusters }
\author{ROBERTO CAPUZZO--DOLCETTA$^1$}
\institute{$^1$Institute of Astronomy, University La Sapienza,
Roma, Italy}
\date{} 
\end{opening}

\begin{document}

\oddpagefooter{}{}{} 
\evenpagefooter{}{}{} 
\bigskip

\begin{abstract}
This short report is concerned with
the well known and not yet satisfactorily answered problem of the existence
of two well distinct typical mass scales  of primordial halo objects: solar size
objects  (halo field stars) and globular cluster size objects ($10^5$ solar masses
and more).  A likely possibility is that almost all the  gaseous content  of the  halo 
fragmented into  massive clouds, which,  in their turn, recycled part of their
gas to the environment in either  a quiescent either a violent way.
The modes and quantity of mass loss depends on thermodynamic properties
of gas clouds, determined, in a zero-metal environment, mainly by 
hydrogenic components evolution and influence on the equation of state.
It is logical to expect that, due to the different dynamical and 
thermodynamical conditions of the recycled gas, another typical fragmenting 
mass scale is settled.
\end{abstract}

\section{Introduction}
Globular clusters constitute a relevant part of the halo of a galaxy, nevertheless
they contain less than 1\% of its mass.  For example, in our Galaxy 
the  average mass of globulars is $\sim 3\times 10^5$ M$\odot$, 
which means a total cluster mass of about $4.5 \times 10^7$  M$_\odot$, whlle the spheroid
mass is about $10^{10}$ M$\odot$,  It is known that globular 
clusters are among the oldest objects in galaxies, aging around $15$ Gyr. 
In spite of the lack of direct determination of the age of 
bulge's stars, it is, nevertheless, commonly accepted for the halo field 
stars an age comparable to that of globulars, inferred also from their low 
metal content.
\par\noindent A diffuse paradigma is that the two components of
the galactic halo (individual stars and globular clusters) are {\it almost 
co--eval}, in the sense that the difference in the formation epoch was 
probably just a fraction of the protogalactic halo free--fall time, which is a 
relatively short time (of the order of a twentieth of the clusters' age).
This paradigma is supported by the evidence of the similar space distribution 
of globular clusters and halo stars, even if these latter seem to be more centrally
peaked than globulars.
\par The question whether or not field stars and globular clusters are coeval 
is out of the purposes of this paper; if we assume that it is logical to
 ask: how can the typical fragmenting mass
change for 5 order of magnitudes in the short time interval that 
is required to allow both
solar mass size objects and globular cluster to fill almost the same 
volume (the halo)?
\par\noindent The use of the Occam's razor make me think that the scientifically most 
{\it economic} point of view is that almost all the 
available gas  fragmented during the proto--halo collapse  into masses of just 
one typical  size  (of either stellar  {\it or} globular cluster 
scale) and later, shortly later, a merging or a  sub--fragmentation lead 
to the other condensed phase. I don't believe as very plausible 
the bottom--up picture (stars formed earlier and partly merged  to 
globulars); much more 
physically viable is, instead, the idea that large clouds condensed first 
and released later
 part of their mass  to the environment, which, consequently, is replenished
of some material available to fragment furtherly on another scale.

\section{A simple picture of halo's structures formation}
In recent years many theories of globular cluster formation have been presented 
in the literature. Most of them are developments of a Fall and Rees (1985) idea.
Fall and Rees  semi--quantitatively supported 
the framework of cluster formation in a two--phase gaseous system: a 
hot ($10^6$ K), dilute gas where clumps are present which are not able to cool 
below $10^4$ K. In this scheme spheroid stars
formed earlier.  
\par\noindent It is interesting to note how various authors, starting from
almost the same picture of Fall and Rees, reached very different, sometimes
contradictory, conclusions. 
 Palla and Zinnecker (1987) found, in a simplified picture, that 
non--equilibrium H2
cooling allows the clumps to cool down to about $100$ K, so that the 
Jeans' mass
drops to stellar values within the clumps. Even if this could explain star 
formation $within$ clumps, i.e. within protoglobular clusters, this does
not explain the $field$ (bulge) star formation.
\par\noindent Ashman (1990) confirms the possibility for the gas to cool to 
$100$ K, but he found that  dark clusters (composed by jupiters) are formed 
from low mass clouds as well as globular clusters from high mass clouds.
Some more detailed scheme and deeper calculations by Murray and Lin (1989) 
show that large amplitude perturbations allow cooling and Pop. III
star formation, whose flux delays globular cluster formation.
Another, more recent, paper by Vietri and Pesce (1995) gave hints to
the idea that imploding shocks may leave stars behind the front.

To summarize, very different conclusions are drawn, and so the question
of how globular clusters and spheroid stars have been formed is actually
still open.

  In this report, we prefer to start, rather than with the Fall and Rees
two--phase gas, with the generally accepted scheme of
a halo which is at its virial temperature, in the range $10^5$-$10^6$ K, so that 
the actual fragmenting mass, of the scale of the Jeans' mass, is much larger
($> 10^7$ M$_\odot$) than the typical globular cluster mass.

These large gaseous clouds collapsed in a more or less violent way 
in dependence on how far from equilibrium they were. This collapse, with its
implications on the cloud thermodynamics and thus on the  equation of 
state of the gas, results in a violent or gentle bounce of the structure.
   In both cases, the existence of a tidal cut-off necessarily implies
that part of the gas mass is lost through the tidal radius to the environment,
on a time scale of the order of the cloud free-fall time.
This way, the halo can be replenished with "processed" gas (shocked or not), 
which is, in any case, likely  in a quite different thermodynamical state
than before.
This would naturally give a diffuse halo at a density and temperature
$significantly$ different from that in the original big clouds, meaning a
$significantly$ different Jeans' mass.

\section{Violent and quiescent mass loss}
The role of the EOS in the evolution  of a self--gravitating gas cloud is 
important, due to the mutual feedback between
dynamics and thermodynamics of the cloud. It is well known, indeed, 
that a {\it soft} EOS induces a much more violent collapse  than a 
{\it harder} EOS in an unstable cloud. The writing of the EOS in the 
usual form

\begin{equation}
p=\left({\gamma -1}\right)\rho u
\end{equation}

(where $p$ is the gas pressure, $\rho$ is the mass density and $u$
is the internal energy per unit mass)
corresponds to a definition of the adiabatic exponent $\gamma$ as
\begin{equation}
\gamma={2\over 3}\left({ 1-{u_{nt}\over u}}\right)+1,
\end{equation}
 $u_{nt}$ being the non--translational part of the internal energy.
\par\noindent Of course, softer EOSs correspond to situations where 
$u_{nt}$ dominates the internal energy (dissociations, ioniziations): 
in self--gravitating
clouds this means that the work of gravitation during a collapse goes
into breaking molecular and/or ionic links among components and not
into the thermal motion which would help the structure to resist the collapse.
  It is quite possible to have clouds in condition
of apparent equilibrium, in the sense that the gas appears almost virialized
but the equilibrium is very unstable because the gas density 
and temperature are such that just a small compressional perturbation makes
$\gamma$ to decrease abruptly inducing the overall structure to fall
under its own gravitation. A possible fate is that of a violent collapse,
which induces high  densities and temperature to be reached in the central
regions, eventually hardening significantly the EOS at a level that 
matter in the centre of the clouds becomes a sort of hard core against which 
the infalling outer layers 'splash' and bounce back at supersonic speed, in 
form of shock waves which carry out of the system a non--negligible 
part of the initial mass of the cloud.

\begin{figure}
\epsfysize=10.0cm 
\hspace{2.5cm}\epsfbox{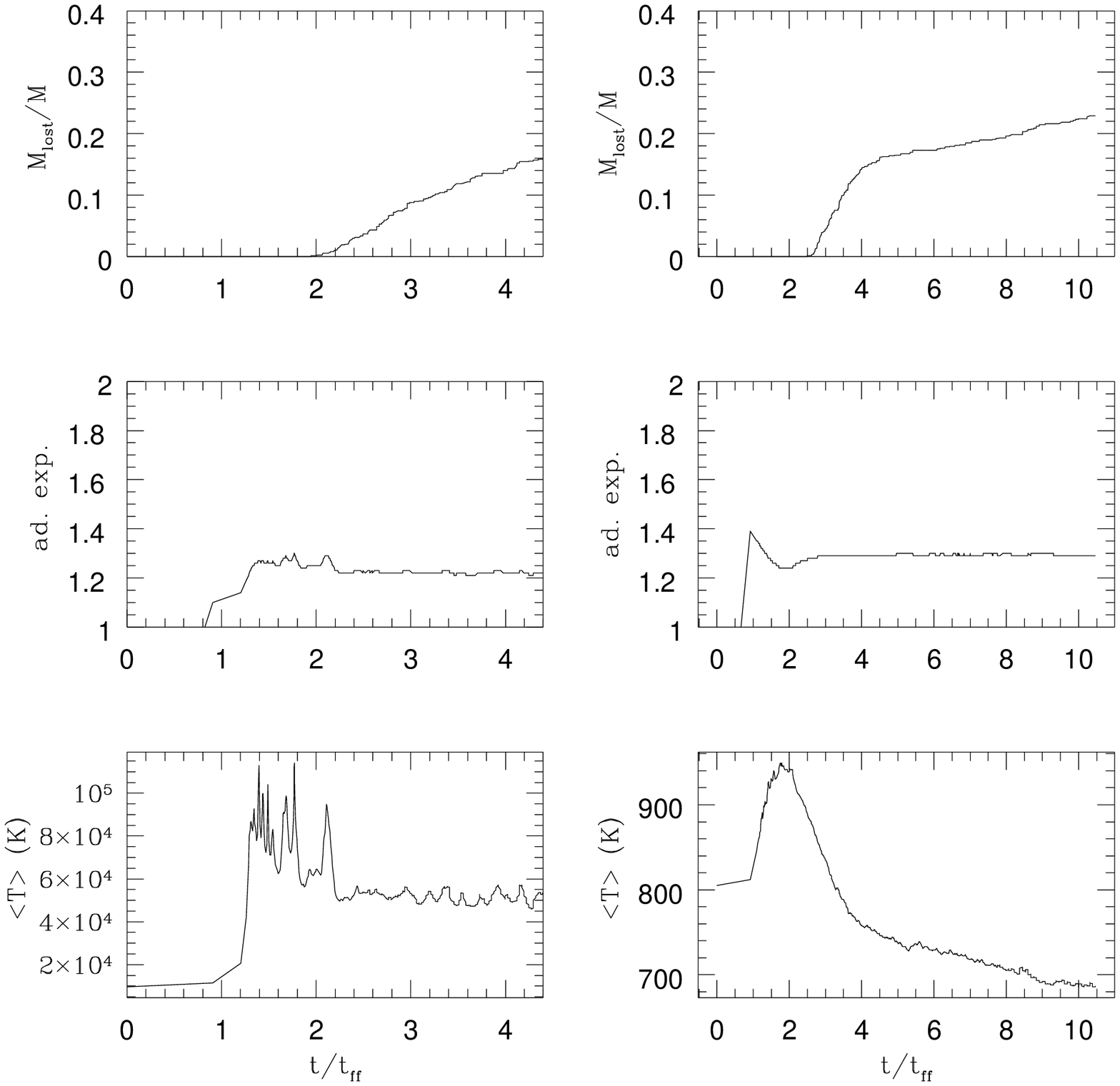}
\caption{} {Evolution of  the average temperature, adiabatic exponent and 
ratio of the mass lost to the total for a $10^8$ M$_\odot$
initially uniform cloud. 
Left panels refer to a cloud with $T_0=10^4$ K; right panels to $T_0=800$ K.
Time is in unit of free--fall time.}
\end{figure}

This is what we call violent mass--loss from protoclouds. 
\par\noindent A careful investigation of such a phenomenon implies 
the development of very detailed and carefully controlled hydrodynamic 
models, for it is known how 
deep collapses of self--gravitating structures are very difficult to
be numerically followed because, with acceptable time steps, 
computational error becomes exceedingly large so to make the
further dynamical evolution  absolutely unreliable.
\par  Note that the numerical error of integration induces quite naturally a 
large error in the total energy, which explains why it is 
possible to have  non-realistic 'explosions' of the whole structure.
\par\noindent An explanation of this can be given through the following simple example.

   It is easily seen that a model of cloud collapse represented by
the equation

\begin{equation}
\ddot R= {p\over \rho} - {GM\over R^2}
\end{equation}

(where $\ddot R(t)$ is the second time derivative of
the radius of the uniformly, $\rho=\rho(t)$, collapsing 
gaseous sphere, $p$ is the pressure at the boundary and $M$ the mass of the 
cloud) leads to unavoidable infinite collapse when the adiabatic exponent
$\gamma$ in the EOS is less than $4/3$. Of course, in real situations
$\gamma$ varies during the cloud evolution, and this has crucial consequences
on the energetic balance.
   Actually, the relative error in the total energy, $\Delta E/ E_0$, is contributed by 
a kinetic term which scales as $-1/R^2$, a gravitational term
proportional to $1/R^2$, and an internal--energy error term scaling
as $R^{ 2-3\gamma}$. 
 The coefficient of $1/R^2$ in the gravitational
error term is estimated to be greater than the kinetic one,
so it is  clear that $\gamma< 4/3$ makes the energy to explode $positively$
as $R$ goes to zero, thus reverting the (correct) extreme collapse
to an eventual explosion whenever  the error in $R$ is not kept 
very well under
control. Actually, if the error in $R$ has grown too much,
it happens  that even if  $\gamma$ re-increases
above the critical $4/3$ the energy may
have definitely changed sign from negative to positive, implying 
unboundedness of the system through an eventual explosion, while
the correct re-expansion, if any, would have been very different.

   The study of violent collapses, and their physical consequences,
requires, so, a very careful attention. I postpone a deep discussion
of this to future work, limiting here to the report of some results 
related to the much easier controlled {\it quiescent} mass loss.
   The quiescent mass loss is what expected when the gas cloud experiences
a gentle contraction and re-expansion and part of the mass of the cloud
escapes through the tidal radius. I have made some simulations
of collapses of self--gravitating gas of primordial compositions, using 
our own SPH code (Capuzzo Dolcetta \& Di Lisio, 1994). Of course the 
adiabatic exponent is allowed to vary,
and its evolution is determined by the evolution of the chemical species
constituting the gas:

$$H,H^+,H^-,H_2,He,He^+,He^{++},e^-.$$

  I don't go  here into details of the models, it suffices to say that
the time evolution of the chemical abundances is followed with a
sophisticated implicit method which is able to reduce the error in their
evaluation and  their feedback on the overall evolution to a very low value.
  Molecular and atomic cooling have been included in the energy
equation.
  
   As an example of the  results, the time evolution of some characteristic 
quantities is shown in 
Figure 1, for two spherical and initially uniform (both in density and
temperature) gas clouds of same mass $M=10^8$ M$_\odot$
and initial radius $R_0=200$ pc, but different initial virial ratios
$Q_0$, namely $Q_0=0.27$ and $Q_0=0.06$ ($hot$ and  $cold$ clouds).  
The clouds are supposed orbiting around the centre of the mother
galaxy at a distance such that the tidal radius is initially $r_t=400$ pc.
\par In both the considered cases the collapse is deepest at about $1.2$ free-fall
times, and the re--expansion leads to a loss of about $20 \%$ of the
total mass in a time shorter than $5t_{ff}$ ($t_{ff}=4$ Myr).
   Note that, whether the characteristic fragmenting mass from the
virial gaseous halo has been 
of the order of $10^8$ M$_\odot$, a loss of $2\times 10^7$ M$_\odot$
{\it per} cloud means that $500$ such clouds may have recycled
a quantity of material of the order of the presently estimated bulge
mass ($\sim 10^{10}$ M$_\odot$).

\section{Conclusions}
It has been shown how delicate is following the evolution of a self--gravitating 
zero--metal cloud in primordial conditions, due to the variation of 
the EOS as consequence of dissociations and inonization of hydrogenic
molecules and ions.
\par As preliminary result of a more extensive future
work, I showed that the $quiescent$ mass loss from protoclouds of
the size  or greater than the Jeans' mass in a virialized protohalo can
provide a substantial fraction of the mass later condensed into stars
of the spheroidal bulge. 

\end{document}